\documentclass[aps,prl,twocolumn,amsfonts,showpacs,superscriptaddress]{revtex4} 
\usepackage{bm}
\usepackage{epsfig,amsopn}
\usepackage{graphicx}
\usepackage{amsmath,amssymb}
\usepackage{natbib}
\def\hM{{\bm{M}}}
\def\hK{{\bm{K}}}
\def\hg{{\boldsymbol{\gamma}}}
\def\hG{{\boldsymbol{\Gamma}}}
\def\bS{{\boldsymbol{\Sigma}}}
\def\bG{{\bm{G}}}
\def\beq{\begin{eqnarray}}
\def\eeq{\end{eqnarray}}
\def\etal{et al.}
\def\nn{\nonumber}
\def\om{\omega}
\def\tX{\tilde{X}}
\def\tV{\tilde{V}}
\def\tv{\tilde{v}}
\def\teta{\tilde{\eta}}
\begin{document}

\title{Generating Function Formula of Heat Transfer in  Harmonic Networks}
\author{Keiji Saito}
\affiliation{Graduate School of Science, 
University of Tokyo, 113-0033, Japan} 
\affiliation{CREST, Japan Science and Technology (JST), Saitama, 332-0012, Japan}
\author{Abhishek Dhar}
\affiliation{Raman Research Institute, Bangalore 560080, India}
\date{\today}

\begin{abstract}
We consider heat transfer across an arbitrary classical harmonic network connected to two
heat baths at different temperatures. The  network has $N$ positional
degrees of freedom, of which $N_L$ are   connected to  a
bath at temperature $T_L$ and $N_R$ are connected to a bath
at temperature $T_R$. We derive an exact formula for the
cumulant generating function  for heat transfer between the two
baths. The formula is valid even for $N_L \ne N_R$ and satisfies the
Gallavotti-Cohen fluctuation symmetry. Since harmonic crystals in
three dimensions are 
known to exhibit different regimes of transport such as ballistic,
anomalous and diffusive, our result implies validity of the fluctuation theorem in all regimes~. 
Our exact formula  provides a powerful tool to study other 
properties of nonequilibrium current fluctuations~.
\end{abstract}

\maketitle 
\section{Introduction}
Nonequilibrium systems typically generate currents of mass or energy.
Understanding  the general features of 
currents and their fluctuations is one of the main goals in nonequilibrium 
statistical physics. 
In this context the fluctuation theorem (FT) is a remarkable discovery 
~\cite{evan1,gc}~. 
This has been theoretically {\cite{evan1,gc,lebo99,Kur98} and experimentally
demonstrated \cite{wang} in many systems.
These studies have pointed to the significance of the large deviation
function (LDF)
of the current
and the related cumulant generating function (CGF) in understanding 
nonequilibrium steady states. 
Some other interesting related developments include the study of  
Bertini and co-workers \cite{BSGJL02}  who introduced a hydrodynamic
fluctuation theory to 
study large dynamic fluctuations in steady states and that of 
Bodineau and Derrida  \cite{BD04,D07}  who conjectured an additivity
principle for 
the LDF and CGF of current, from which one can predict the
quantitative behavior of higher order correlations of currents.

In the context of transport, most analytic results on 
LDFs and FTs are on systems where the bulk dynamics is
stochastic, such as the simple exclusion process,  zero range process,
Brownian motors, etc.  \cite{lebo99,D07,lacoste08}. 
It is of general interest to consider and develop these arguments for
systems with bulk Hamiltonian dynamics.
However, exact analysis here is generally difficult. For
example even the problem of demonstrating Fourier's law of heat conduction in
a deterministic system has proved to be difficult and has led to many
surprises \cite{BLR00,LLP03,dhar08}. Heat conduction in harmonic
lattices is one 
exception  where many  nonequilibrium properties can be precisely
discussed. 
The average heat current, the main focus of work so far,  is   given by a
Landauer-like formula  in terms of phonon transmission coefficients
\cite{dharroy06}.  
Using the transport formula it was recently demonstrated  \cite{kundu10}
numerically that  disordered harmonic crystals in two 
and three dimensions can exhibit different regimes of transport such as
ballistic, localized, anomalous and diffusive.   
Given that this  simple deterministic model exhibits various regimes of
transport, it is of
interest to study the generic features of nonequilibrium current
fluctuations in this system. 

In this paper, we derive the general formula of the CGF for heat
current in an arbitrary harmonic lattice connected to two heat baths, which 
provides a basis to explore generic features of current fluctuations. 
Consider heat transfer through a system from a bath at temperature
$T_L$ to a bath at temperature $T_R$.
Let $Q$ be the heat transferred from the left reservoir to the system during
measurement time $\tau$. 
In general, the distribution of heat $P(Q)$ has an asymptotic form $P(Q)\sim
e^{\tau h(q) }$ at large $\tau$
where ${h}(q=Q/\tau)$ is the LDF. 
The CGF $\mu(\lambda)$  generates cumulants of the heat transfered and is
defined through the relation: 
$\langle e^{\lambda Q } \rangle \sim e^{\tau \mu (\lambda
  )}$. The LDF $h(q)$ and the CGF $\mu(\lambda )$  are connected through the
Legendre transform $\mu(\lambda ) = \max_{q} \left[ \lambda q + h(q) \right]~$. 
Properties of heat current fluctuations are contained in  $h(q)$ or
equivalently in $\mu(\lambda)$ and various results    such as the 
steady state FT and the additivity principle conjecture  
can be  stated in the framework of either the LDF or the CGF. 
For heat conduction, the steady state FT of Gallavotti-Cohen (GC) \cite{gc}
implies the symmetry relation $\mu(\lambda)=\mu(-\lambda-\Delta \beta)$, where
$\Delta \beta=1/T_R-1/T_L$, and is referred to as the GC symmetry.  
There are examples where the symmetry of $\mu(\lambda)$
does not imply the FT  \cite{ftviol}~.  
However, the CGF and its symmetry
property  themselves provide important information on current fluctuations. 
Among these, one of the most
interesting consequences of the symmetry relation is  that it 
leads to the standard linear response results such as
Onsager reciprocity and Green-Kubo relations \cite{GG96,lebo99}, 
and in addition  makes predictions of responses 
in the far from equilibrium regime\cite{SU08,AG07,K10}.
So far, for Hamiltonian systems, the CGF has been analytically obtained for
one and two particle  systems \cite{visco06,wijland06} and for a one-dimensional quantum harmonic chain
\cite{SD07}~.  
Here we obtain a general  formula for the CGF of a harmonic system in terms of
the transmission matrix of a phonon mode $\omega$ from one  reservoir to 
the other. Remarkably, the expression is robust regardless of the complexity
of the network and the number of particles that 
are attached to reservoirs and  always satisfies the GC symmetry.

\section{Harmonic Networks and Heat Transfer} 
\begin{figure}
\includegraphics[width=6.5cm]{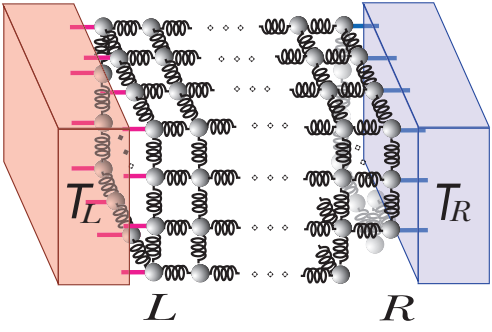}
\caption{
(color online). A schematic picture of the harmonic network connected to reservoirs.
Positional degrees of freedom is labelled by $i=1,\cdots , N$. 
The points at $\ell = i \in L$ and $r = i \in R$ 
are attached to the left and right reservoirs. }
\label{fig1}
\end{figure}  
We consider an arbitrary classical harmonic network with $N$ positional degrees
of freedom labelled 
$i=1,2,\cdots,N$ of which  $N_L$ are connected 
to a bath at temperature $T_L$ and $N_R$ are  connected to a bath at
temperature $T_R$ (see figure $1$). 
As an example, for a three-dimensional cubic crystal consisting of $N_c^3$
atoms with vector displacements, and with two opposite faces coupled
to heat baths, we would  have $N=3N_c^3$ and $N_L=N_R=3 N_c^2$~. 
We denote the positional degrees and their corresponding
velocities by the column vectors $X=(x_1,x_2,\cdots,x_N)^T$ and
$V=(v_1,v_2,\cdots,v_N)^T$ respectively. We consider the following
general harmonic Hamiltonian for the system:
\beq
H=\frac{1}{2} V^T \hM V+\frac{1}{2} X^T \hK X ~, \label{ham}
\eeq 
where $\hM={\rm diag}\{m_1,m_2,\cdots,m_N\}$ denotes the mass matrix and
$\hK$ the force matrix for the system. 
We model the heat baths by white noise Langevin equations with each
variable coupled to a bath having an independent Langevin
dynamics. 
Let $L~ (R)$ refer to the set of $N_L (N_R)$ points connected
to the left~(right) bath.
For discriminating these points from the bulk points, 
we use indices $\ell=i \in L$ and $r=i \in R$~.
We define the $N$-component noise vectors $\eta = \eta^{(L)} + \eta^{(R)}$ such
that $\eta_{\ell}=\eta^{(L)}_{\ell}$ and $\eta_r=\eta^{(R)}_r$ are non-zero.
Also we define the diagonal matrices $\hg = \hg^{(L)} + \hg^{(R)}$
such 
that $\hg_{\ell \ell}=\hg^{(L)}_{\ell \ell}\equiv \gamma_\ell$ and 
$\hg_{rr}=\hg^{(R)}_{rr}\equiv \gamma_r$ are nonzero.
The equations of motion for the system are then given by:
\beq
\dot{X}&=&V \nn \\
\hM \dot{V}&=& -\hK X -\hg V +\eta   \nn  \\
&=& -\hK X -\hg^{(L)} V +\eta^{(L)}~-\hg^{(R)} V+\eta^{(R)}~. \label{eqm}
\eeq 
The noise terms are assumed to be Gaussian white noise with zero mean and correlations given by 
$
\langle \eta_{\ell} (t) \eta_{\ell '} (t') \rangle = 2 \gamma_\ell T_L \, \delta_{\ell , \ell '}
\delta(t-t' ) 
$, 
$\langle \eta_{r} (t) \eta_{r '} (t') \rangle = 2 \gamma_r T_R \, \delta_{r , r '}
\delta(t-t' )$, and $\langle \eta_{\ell} (t) \eta_{r} (t') \rangle = 0 $,
where we have set the Boltzmann constant to the value one.
The initial state at $t=0$ is chosen from the steady
state distribution, and we measure the heat $Q$ flowing from the left
reservoir  into the system between the times 
$t=0$ to $t=\tau$~. We thus have:
\beq
Q=\sum_\ell \int_0^\tau dt~ v_\ell~ (-\gamma_\ell v_\ell + \eta_\ell)~. \label{heat}
\eeq 
A solution of the linear equations Eq.~(\ref{eqm}) can be obtained by
introducing the following discrete Fourier transforms and their inverses: 
\beq
\{X(t),V(t), \eta(t)\} = \sum_{n=-\infty}^\infty
\{{\tX}(\omega_n),{\tV}(\omega_n), \teta(\omega_n) \} e^{-i \omega_n t} ~ , \nn \\
\{\tX(\omega_n),\tV(\omega_n), \teta(\omega_n)\} = 
\frac{1}{\tau}
 \int_0^\tau 
\{{X}(t),{V}(t), \eta(t) \} ~e^{i \omega_n t} \nn ~,
\eeq 
where $\omega_n = {2\pi n / \tau}$. 
Plugging these into Eq.~(\ref{eqm}), we get:
\beq
\tV(\omega_n)&=& -i \omega_n {\bm G}^+(\omega_n)  ~[\teta^{(L)}(\omega_n)
  +\teta^{(R)}(\omega_n)] \nn \\ 
&+&\frac{1}{\tau} \bG^+(\omega_n)~ [~\hK \Delta X + i \omega_n \hM \Delta V ~]~, \label{gexp} \\
{\bm G}^+(\omega_n)& =& \left[ -{\bm M} \omega_n^2 +{\bm K} - 
  {\bS^{(L)}} (\omega_n ) -\bS^{(R)} (\omega_n ) \right]^{-1} \!\!, ~~~~\label{gf}
\eeq
where $\bS^{(L,R)}(\omega)= i \omega {\hg^{(L,R)}}$~, 
$\Delta X= X(\tau)- X(0),$ and $\Delta V= V(\tau)-V(0)$. 
The matrix $\bG^+$ is the Green's function connecting bulk variables with  
reservoir properties. 
The noise correlations in the Fourier space are given by: 
\beq
\begin{array}{c}
\langle {\teta}_{\ell} (\omega_n ) {\teta}_{\ell '} ( \omega_{n'} ) \rangle 
= 2 \delta_{\ell , \ell '}  \delta_{n,-n'} \gamma_\ell T_L  \, / \tau  \, , \\
\quad  \label{noise_omega} \\
\langle {\teta}_{r} (\omega_n ) {\teta}_{r '} ( \omega_{n'} ) \rangle 
= 2 \delta_{r , r '}  \delta_{n,-n'} \gamma_r T_R \, / \tau .
\end{array}
\eeq 
Since the noise strength $\teta(\omega_n) \sim O(1/\tau^{1/2})$ and $\Delta
X, \Delta V \sim O(1)$ we see that the second term in Eq.~(\ref{gexp}) 
is $\sim 1/\tau^{1/2}$  order smaller than the first and so can be
dropped. It can in fact be shown that it contributes 
order $1/\tau$ corrections to the CGF \cite{kundu10v2}. 
We note that 
$\tV^*(\omega_n )=\tV(-\omega_n),~\teta^*(\omega_n)=\teta(-\omega_n)
$~. The heat transfered $Q$  [Eq.~(\ref{heat})] can be expressed 
in terms of the Fourier-modes with $n \geq 0$ as 
$Q=\sum_{n=0}^\infty\sum_{\ell}  
[-2 \gamma_{\ell} \tv_{\ell}(\omega_n) \tv^*_{\ell} (\omega_n) +
\teta_\ell(\om_n) \tv^*_\ell (\om_n)+ \tv_\ell(\om_n) \teta_\ell^*(\om_n)]. $
We define $\bG^-(\omega )=\bG^+(-\omega )=[{\bG^+}(\omega ) ]^*$
and $\hG^{(L,R)} (\omega ) = {\rm Im}\{\bS^{(L,R)}(\omega ) \} = \omega \hg^{(L,R)}$. 
On using the solution (\ref{gexp}) without the second term, i.e.,
$\tV(\om_n)=-i \omega_n\,\bG^+ (\om_n)\, [\teta^{(L)}(\om_n)
+\teta^{(R)} (\om_n) ]$, we get the expression of $Q$ as:
\begin{widetext}
\beq
Q 
&=&  
\tau~ \sum_{n=0}^{\infty}  
 ( -2 \omega_n ) \big[
{\teta}_{\ell '} (\omega_n )  {\bG^+}_{\ell ' \ell} 
+ 
{ \teta}_{r } (\omega_n ) 
{\bG^+}_{r \ell}
\big]  \hG^{(L)}_{\ell \ell} \, 
\big[  
{\bG}_{\ell \ell''}^-
 { \teta}^*_{\ell '' }(\omega_n )
+   {\bG}_{\ell r'}^- 
 { \teta}^*_{r'} (\omega_n )  
\big]  \nonumber \\
&&- i \omega_n 
 \big[
{\teta}_{\ell '} (\omega_n )  ({\bm G}^+_{\ell ' \ell} -{\bm
  G}^-_{\ell ' \ell})  
+ {\teta}_r (\omega_n ) 
({\bm G}^+_{r \ell}-{\bm G}^-_{r \ell})
\big]~ \teta^*_\ell(\om_n)  \label{Q-expression}\\
 &=&\tau~\sum_{n=0}^{\infty}
(\tilde{\bm \eta}_L (\omega_n ) ,\tilde{\bm \eta}_R (\omega_n ) ) \, {\cal A}  \, 
\left(
\begin{array}{c}
\tilde{\bm \eta}^*_L ( \omega_n ) \\
\tilde{\bm \eta}^*_R ( \omega_n ) 
\end{array} 
\right) , \nn 
\eeq
\end{widetext}
where 
$\tilde{\bm \eta}_L$ ($\tilde{\bm \eta}_R$) denotes  an $N_L$ ($N_R$) component
column vector  of noise belonging to $\ell \in L$ ($r \in R$) sites,
while the  $N_L+N_R$ dimensional hermitian matrix ${\cal A}$ is given by
\begin{widetext}
\beq
{\cal A} &=& 
\left( 
\begin{array}{ll}
2 \omega_n [{\bm G}^+ {\bm \Gamma}^{(R)} {\bm G}^- ]_{LL}  \,  & 
i\omega_n [{\bm G}^-]_{LR} -2\omega_n [{\bm G}^+ {\bm \Gamma}^{(L)} {\bm G}^- ]_{LR}  \\
-i\omega_n  [{\bm G}^+]_{RL} - 2\omega_n 
[{\bm G}^+{\bm \Gamma}^{(L)} {\bm G}^-]_{RL}  & -2\omega_n 
[{\bm G}^+{\bm \Gamma}^{(L)} {\bm G}^- ]_{RR}
\end{array}
\right) ~ .    \label{a-matrix}
\eeq
\end{widetext}
The subscript $L$ and $R$ in the matrices respectively represent the space 
of $\ell$ and $r$-sites. In Eqs.(\ref{Q-expression}) and (\ref{a-matrix}),
the $\omega_n$ dependence of ${\bG}$ and $\hG$ 
have been suppressed. In what follows, the $\omega_n$ dependence in variables 
is omitted unless it is necessary.
In obtaining the $(L,L)$ element of the matrix ${\cal A}$, we have used the
following Green's function identity which easily follows from the definition
(\ref{gf}):   
\beq
{\bm G}^+  -{\bm G}^- &=& 2i {\bm G}^- 
\,(\hG^{(L)} +\hG^{(R)} )\, {\bm G}^+  \nn \\ 
&=&  2i {\bm G}^+ \,(\hG^{(L)}+\hG^{(R)})\,  {\bm G}^-. 
~~~~\label{grelation}
\eeq
\section{Cumulant Generating Function}
We now proceed to the calculation of the CGF. 
The characteristic function ${\cal Z}(\lambda)=\langle e^{\lambda Q} \rangle$ is obtained by using  the expression in
Eq.~(\ref{Q-expression}) and averaging over the Gaussian noise
variables whose correlation matrix is given in Eq.~(\ref{noise_omega}). We
get: 
\beq
{\cal Z}(\lambda)
&=& 
{\cal N}
\prod_{n \ge 0}\int d[
\tilde{\bm \eta}_L ,\tilde{\bm \eta}_R , \tilde{\bm \eta}_L^\ast ,\tilde{\bm \eta}_R^\ast 
] 
\exp \Bigl[ 
~\tau \, (\tilde{\bm \eta}_L , \tilde{\bm \eta}_R ) 
\nonumber \\
&\times&\!\!\! 
\left[ \lambda {\cal A} 
- \left(
\begin{array}{cc}
{1 \over 2 T_L }[\hG^{(L)}]^{-1} 
  & 0 \\
0   &  {1 \over 2  T_R }[\hG^{(R)}]^{-1} 
\end{array}
\right)
\right]\!
\left( 
\begin{array}{c}
\tilde{\bm \eta}_L^* \\ \tilde{\bm \eta}_R^*
\end{array}
\right)
\Bigr]  , \nonumber  \\
\eeq
where
${\cal N}$
denotes the normalization factor of the noise distribution.
By performing the Gaussian integral, one obtains the formal 
expression of the CGF:
\beq
\mu(\lambda ) &=& {1\over \tau} \log {\cal Z} (\lambda ) \Bigr|_{\tau\to\infty} =
-{1\over \tau}\sum_{n\ge 0} \log \det {\cal B}\Bigr|_{\tau\to\infty}  , \label{detBB}~~~~~~~\\
{\cal B} 
&=& {\bm 1} - \lambda 
\left( 
\begin{array}{cc}
{\cal T}_L & {\cal T}_{LR} \\
{\cal T}_{RL} & - {\cal T}_R  \\
\end{array}
\right)
\left( 
\begin{array}{cc}
T_L {\bm 1} &  {\bm 0}  \\
{\bm 0} & T_R{\bm 1}  \\
\end{array}
\right) , 
\eeq
where
\beq
{\cal T}_L &=& 4 [{\bm G}^+ {\bm \Gamma}^{(R)} {\bm G}^- {\bm \Gamma}^{(L)}]_{LL} , \nn \\
{\cal T}_R &=& 4 [{\bm G}^+ {\bm \Gamma}^{(L)} {\bm G}^- {\bm \Gamma}^{(R)}]_{RR} , \nn \\
{\cal T}_{LR} &=& 
-4 [{\bm G}^+ {\bm \Gamma}^{(R)} {\bm G}^- {\bm \Gamma}^{(R)}]_{LR} 
+2i[{\bm G}^- {\bm \Gamma}^{(R)}]_{LR}, \nn \\
{\cal T}_{RL} &=& 
-4 [{\bm G}^+ {\bm \Gamma}^{(L)} {\bm G}^- {\bm \Gamma}^{(L)}]_{RL} 
-2i[{\bm G}^+ {\bm \Gamma}^{(L)}]_{RL} . \nn 
\eeq
The matrices ${\cal T}_L$ and ${\cal T}_R$ are respectively 
$N_L\times N_L$ and $N_R \times N_R$ square matrices, and these 
can be regarded as transmission amplitude of energy 
with the mode $\omega_n$ from one reservoir to the other. 
For $N_L=N_R$, it is known that these matrices appear in the 
Landauer-like formula for average current \cite{dharroy06}.
As clarified later, even for unequal case $N_L \ne N_R$, both of these
are transmission matrices and enter in the Landauer-like formula.
Although the physical meaning of ${\cal T}_{LR}$ and ${\cal T}_{RL}$ are
not clear,  these are closely related to ${\cal T}_L$ and ${\cal T}_R$.
The relations can be revealed by using the relation 
(\ref{grelation}) iteratively. Through tedious but straightforward calculations,
one finds the following nontrivial relations:
\beq
{\cal T}_{LR} {\cal T}_R &=& {\cal T}_L {\cal T}_{LR} , \label{rel1}\\
{\cal T}_{RL} {\cal T}_L &=& {\cal T}_R {\cal T}_{RL} , \label{rel2}\\
{\cal T}_{RL} {\cal T}_{LR} &=& {\cal T}_R ( {\bm 1} - {\cal T}_R ) . \label{rel3}  
\eeq  
In order to get simple form of the CGF, we need to simplify the determinant of 
${\cal B}$ in Eq.~(\ref{detBB}). 
The relations (\ref{rel1})-(\ref{rel3}) play a central role in
simplifying the determinant of ${\cal B}$ and in 
deriving the final expression of the CGF. 
We heuristically introduce the matrix ${\cal C}$:
\beq
{\cal C} &=& 
\left( 
\begin{array}{cc}
{\bm 1} & {\cal C}_{LR} \\
{\bm 0} & {\cal C}_{RR} 
\end{array}
\right) \nonumber \\ 
{\cal C}_{LR} &=& \lambda T_R {\cal T}_{LR} + {T_R \over T_L} {\cal T}_L {\cal T}_{LR} \, \\
{\cal C}_{RR} &=& {\bm 1} + \left( {1\over \lambda T_L} -\lambda T_L - 1\right){\cal T}_R
+{\cal T}_{RL}{\cal T}_{LR} \nonumber \\
&=& \left( {\bm 1} + {{\cal T}_R \over \lambda T_L}\right)
\left( {\bm 1} - \lambda T_L {\cal T}_R \right) . 
\eeq
The advantage of introducing the matrix ${\cal C}$ is that the product ${\cal
  B}{\cal C}$ has a simple form, and this is useful to simplify 
$\det {\cal B}$ given by $\det{\cal B } {\cal C} / \det{\cal C} =\det{\cal B } {\cal C} / \det{\cal C}_{RR} $. 
With the relations (\ref{rel1})-(\ref{rel3}), one finds the following 
form for the product: 
\begin{widetext}
\beq
{\cal B}{\cal C} &=& 
\left( 
\begin{array}{cc}
{\bm 1} -\lambda T_L {\cal T}_L   & {\bm 0} \\
-\lambda T_L {\cal T}_{RL}
 & \left( {\bm 1} + {{\cal T}_R \over \lambda T_L}\right) 
\left( 
{\bm 1} - {\cal T}_R  T_L T_R \lambda (\lambda + \Delta \beta )
\right)
\end{array}
\right)~,
\eeq 
\end{widetext}
and hence: 
\beq
\det {\cal B}= \det \left( 
{\bm 1} - {\cal T}_R  T_L T_R \lambda (\lambda + \Delta \beta )
\right) \frac{\det ({\bm 1} -\lambda T_L {\cal T}_L )} { \det ( {\bm 1} -\lambda
  T_L {\cal T}_R )} .~~\label{detB}
\eeq 
Now by taking the singular value decomposition of the matrix
$[(\hG^{(L)})^{1/2} {\bm G}^+ (\hG^{(R)})^{1/2} ]_{LR} $ it can be
shown that ${\cal T}_L$ and ${\cal T}_R$ have the same set of non-zero
eigenvalues. Hence $ \det ( {\bm 1} -\lambda   T_L {\cal T}_L )
=\det ( {\bm 1} -\lambda   T_L {\cal T}_R ) $ and on using this in  
Eq.~(\ref{detB}) we get, in the large $\tau$ limit:
\beq
\mu(\lambda ) &=& -{1\over 2\pi}
\int_{0}^{\infty} d\omega  \nonumber \\
&&\times ~{\rm Tr}
\log \Bigl[ {\bm 1}
- {\cal T} (\omega ) T_L T_R 
\lambda \left( \lambda + \Delta \beta \right)  \Bigr] ~,~~~~~ \label{formula}
\eeq
where one can use either ${\cal T}_L$ and ${\cal T}_R$ for the 
transmission matrix  ${\cal T}(\omega )$, both of which generate 
the same values of current cumulants.
This formula for the CGF is the central result of this paper.
One can easily check that the GC symmetry:  
$\mu(\lambda ) = \mu(-\lambda -\Delta \beta)$ is satisfied.
When the system is 
one dimensional and $N_L=N_R=1$, the 
formula reproduces the classical limit of the quantum version of CGF
\cite{SD07}. 
Interestingly, the formula (\ref{formula})
is valid even for $N_L \ne N_R$. 
In this paper, for simplicity we demonstrated the derivation for baths
with  white Gaussian noise. 
However, the formula is also valid for 
generalized Langevin noise  with memory kernel, with appropriate
definition  of the matrices ${\bS}$ and ${\bm \Gamma}$.

\section{Discussion}
We have derived an exact formula for the CGF of a general harmonic
network (\ref{formula}) and shown that it satisfies the GC symmetry. 
The formula is expressed in terms  of the phonon transmission matrix.
The CGF can be used to obtain the average current 
$\langle Q \rangle_c /\tau $ and 
current noise $\langle Q^2 \rangle_c /\tau $
by taking the first and second derivatives with respect to $\lambda$:
\beq
{\langle Q \rangle_c \over \tau} 
&=& {(T_L - T_R ) \over 2\pi }
\int_{0}^{\infty} d \omega \, {\rm Tr}  [{\cal T} (\omega )]\nonumber . \\
{\langle Q^2 \rangle_c \over \tau} 
&=& 
{1 \over 2\pi }
\int_{0}^{\infty} \!\! d \omega \, 
{\rm Tr} [{\cal T}^2 (\omega ) (T_R - T_L)^2
+2{\cal T} (\omega ) T_L T_R ] \nonumber .
\eeq 
Higher order cumulants are also systematically given.

The transmission matrix can be obtained either analytically or numerically. 
In case of higher dimensional regular lattices, 
the recursive Green's function method \cite{kundu10v2} can be used to  
efficiently generate the transmission matrix and thus evaluate the CGF. 
The disordered harmonic lattice shows different regimes of transport,
such as ballistic,  anomalous and diffusive transports and hence
our result implies validity of the fluctuation theorem in all regimes of transport. 
In addition, the formula (\ref{formula}) can be a powerful tool
to explore generic features of current fluctuation in these different regimes.
One of the most interesting possible application would be a test of the
conjecture  of the additivity principle \cite{BD04}.

An open problem is the quantum expression of the CGF. As in the
one-dimensional case \cite{SD07},  
a two-point observation protocol to get distribution of heat is
necessary, and this seems to be a much more complex calculation than
the one in this paper.  

\section*{Acknowledgments}
We thank the Centre for Computational Science and Engineering,
National University of Singapore where this work was 
initiated. KS was supported by MEXT, Grant Number (21740288).

\end{document}